\documentclass[conference]{IEEEtran}
\IEEEoverridecommandlockouts
\usepackage{cite}
\usepackage{amsmath,amssymb,amsfonts}
\usepackage{algorithmic}
\usepackage{graphicx}
\usepackage{textcomp}
\usepackage{xcolor}
\usepackage[dvipsnames]{xcolor}
\usepackage{multirow}
\usepackage{adjustbox}
\usepackage{makecell}
\usepackage{tabularx}

\usepackage{subfig}
\usepackage{soul}

\usepackage{booktabs}
\usepackage{url}
\usepackage{mdframed}
\usepackage{listings}
\usepackage[most]{tcolorbox}
\usepackage{hyperref}

\def\BibTeX{{\rm B\kern-.05em{\sc i\kern-.025em b}\kern-.08em
    T\kern-.1667em\lower.7ex\hbox{E}\kern-.125emX}}

\newcounter{example}[section]
\renewcommand{\theexample}{\thesection.\arabic{example}}

\newmdenv[
  linecolor=gray,
  linewidth=0.5pt,
  backgroundcolor=gray!10,
  roundcorner=5pt,
  innertopmargin=8pt,
  innerbottommargin=8pt,
  innerleftmargin=10pt,
  innerrightmargin=10pt,
  skipabove=12pt,
  skipbelow=12pt
]{examplebox}

\lstdefinestyle{jsStyle}{
  language=Java,
  backgroundcolor=\color{gray!10},
  basicstyle=\ttfamily\small,
  frame=none,
  breaklines=true,
  showstringspaces=false,
  keywordstyle=\color{blue},
  commentstyle=\color{gray}
}

\tcbuselibrary{breakable}

\begin{document}

%\title{Investigating Software Aging in Automatically Generated Software Systems}
% \title{Investigating Software Aging in LLM-Generated Software Systems}
\title{Software Aging in LLM-Generated Applications: Runtime Evidence, Static Analysis, and Human-Written Comparisons}

\author{
\IEEEauthorblockN{{César Santos}\\
%\IEEEauthorblockA{\textit{Department of Computing} \\
\textit{Federal Rural University of Pernambuco}\\
Recife, Brazil \\
cesar.santos@ufrpe.br}
\and
\IEEEauthorblockN{{Michele Vitagliano}\\
%\IEEEauthorblockA{\textit{Department of Information Technology and Electrical Engineering} \\
\textit{University of Naples Federico II}\\
Naples, Italy \\
mic.vitagliano@studenti.unina.it}
\and
\IEEEauthorblockN{{Roberto Natella}\\
\textit{Gran Sasso Science Institute (GSSI)}\\
L'Aquila, Italy\\
roberto.natella@gssi.it}
\and
\IEEEauthorblockN{{Ermeson Andrade}\\
%\IEEEauthorblockA{\textit{Department of Computing} \\
\textit{Federal Rural University of Pernambuco}\\
Recife, Brazil \\
ermeson.andrade@ufrpe.br}
}

\maketitle

\begin{abstract}

Large Language Models (LLMs) are increasingly used to generate executable software systems from natural language specifications, accelerating development and reducing manual implementation effort. Although recent studies have investigated the functional correctness, security, maintainability, and robustness of LLM-generated code, little is known about the long-term reliability of such systems under sustained execution. 
In this paper, we experimentally investigate software aging symptoms in LLM-generated service-based applications across generation-and-execution environments. Using backend scenarios derived from BaxBench, we generated applications targeting JavaScript, Python, and Rust through LLM-based generation platforms, validated them with BaxBench-derived tests, and subjected them to 48-hour workload executions. We monitored memory usage, response time, and throughput and analyzed them using the Mann--Kendall test and Sen's slope estimator. We further complemented the runtime evaluation with static analysis of the generated source code and an exploratory comparison with human-written implementations of related backend scenarios. The results show that memory usage is the most consistent indicator of potential software aging, with statistically significant upward trends in most application-environment combinations, while response time and throughput exhibit more heterogeneous behavior. Static analysis identified plausible code-level aging mechanisms, and the comparison with human-written systems showed that the aging symptoms observed in LLM-generated applications align with degradation patterns also found in manually developed implementations. These findings indicate that functional correctness alone is insufficient to assess the operational reliability of LLM-generated software before deployment in continuously running environments.
\end{abstract}

\begin{IEEEkeywords}
LLM-generated software, software aging, software reliability, static analysis, human-written software, performance degradation, empirical study.
\end{IEEEkeywords}

\section{Introduction}

Large Language Models (LLMs) have significantly changed the way software systems are designed and implemented. Tools such as GitHub Copilot, ChatGPT, and other LLM-based code generation platforms can produce executable code from natural language descriptions, API specifications, or partial implementation requirements. This capability has accelerated prototyping and development, reducing manual programming effort and enabling developers to generate complete applications from high-level prompts~\cite{2024Accelerating,chauhan2025llm}. As a result, LLM-generated software is increasingly relevant in both research and practical software engineering contexts.

Despite these advances, the reliability of LLM-generated software remains an open concern. Recent studies have investigated several quality attributes of generated code, including functional correctness, security, maintainability, robustness, and performance~\cite{Lyu2024Automatic,Jamil2025Can,Molison2025Is,Kharma2025Security,Shehab2024Evaluating,Tambon2024Bugs,Liu2024Beyond,Peng2025CWEval}. Benchmarks such as BaxBench~\cite{Vero2025BaxBench:} evaluate whether LLMs can generate correct and secure backend applications from natural language specifications. However, most existing evaluations focus on whether the generated software satisfies functional requirements or avoids specific classes of defects. Less attention has been given to the behavior of these systems during prolonged execution, especially when they are deployed as long-running backend services.

This limitation is important because a system that passes functional tests may still degrade over time. In long-running systems, software aging refers to the progressive degradation of performance or dependability during execution, often caused by resource leaks, unreleased memory, fragmentation, accumulated errors, or other runtime effects~\cite{Parnas1994,Grottke2008The,cotroneo:2014}. Software aging has been widely studied in traditional systems, including web servers, databases, cloud and edge environments, blockchain platforms, testing frameworks, and machine learning-based applications~\cite{Grottke2006Analysis,Matias2010Accelerated,pietrantuono2022empirical,SqlServerAgingNascimento2024,couto2024comparative,AgingTestFrameworkDias2025}. Nevertheless, it remains unclear whether applications generated by LLMs exhibit similar aging symptoms when executed continuously under sustained workloads.

Investigating this problem is particularly relevant because LLM-generated applications may contain implementation choices that are functionally correct but inefficient during long-duration execution. For example, generated code may repeatedly allocate resources without adequate cleanup, handle files or network connections inefficiently, or introduce runtime behaviors that only become visible after several hours of execution. These issues may not be detected by conventional functional tests, but they can affect memory consumption, latency, throughput, and ultimately the operational reliability of generated software.

To address this gap, this paper presents an empirical investigation of software aging symptoms in LLM-generated service-based applications across generation-and-execution environments. We use application scenarios derived from BaxBench to generate backend services in JavaScript, Python, and Rust. The JavaScript implementations were generated using Bolt and implemented with Node.js and Express, the Python implementations were generated using ChatGPT 5.0 and implemented with FastAPI, and the Rust implementations were generated using Gemini 3.0 and implemented with Actix Web. All implementations of the same application scenario preserved the same endpoints, route names, input formats, and expected behavior. After functional validation, each application-language combination was subjected to a 48-hour workload execution, during which memory usage, response time, and throughput were monitored. Beyond the runtime experiments, we also analyze the generated source code using traditional static-analysis tools and AI-assisted code review to identify plausible code-level mechanisms that may explain the observed aging symptoms. Finally, we provide an exploratory comparison between the LLM-generated applications and functionally related human-written open-source implementations, in order to contextualize whether the observed degradation trends are specific to generated code or also appear in manually developed systems.

The main contributions of this article are as follows:

\begin{itemize}
\item We extend software aging research to the context of LLM-generated backend applications, a class of systems that has mainly been evaluated in terms of correctness, security, maintainability, and code quality rather than long-term operational reliability.

\item We design and execute a long-duration experimental methodology to evaluate software aging symptoms in LLM-generated applications, using memory usage, response time, and throughput as observable indicators of degradation.

\item We provide a cross-environment empirical analysis covering JavaScript/Node.js/Express, Python/Flask, and Rust/Actix Web implementations generated by different LLM-based platforms. Rather than isolating the effect of programming language, this analysis examines whether aging symptoms emerge across a diverse set of generation and execution environments.

\item We complement the runtime evaluation with a static analysis of the LLM-generated applications, identifying plausible code-level aging mechanisms, such as unbounded persistent state, missing cleanup paths, and conditional resource retention, and examining their correspondence with the degradation trends observed at runtime.

\item We provide an exploratory comparison between LLM-generated and human-written implementations of related backend scenarios, showing that manually developed systems can also exhibit aging trends that are comparable to, or stronger than, those observed in generated applications.

\item We discuss the implications of software aging for the deployment of LLM-generated applications, showing that passing functional validation is not sufficient to assess the operational reliability of generated software intended to run continuously.
\end{itemize}

The remainder of this paper is organized as follows. Section~\ref{AutoGenSoftwareSys} provides background on automatically generated software systems. Section~\ref{relatedWork} reviews related work on LLM-generated software and software aging. Section~\ref{experiments} describes the experimental plan and setup, including the research questions. Section~\ref{results} presents the runtime aging analysis, the cross-environment comparison, the static analysis of the LLM-generated source code, and the exploratory comparison with human-written implementations. Finally, Section~\ref{conclusion} concludes the paper and presents future research directions.

\section{Automatically Generated Software Systems}
\label{AutoGenSoftwareSys}

Automatically generated software refers to software systems produced with limited manual coding effort through automated tools, model transformations, code generators, or artificial intelligence-based techniques. Traditional forms of software generation include model-driven development, API generators based on specifications such as OpenAPI, object-relational mapping frameworks, compilers, transpilers, and low-code/no-code platforms. These approaches automate parts of the development process by translating higher-level descriptions, models, or schemas into executable code or reusable software components, reducing the gap between system specification and implementation~\cite{Lyu2024Automatic,yang2024automated}.

Recent advances in LLMs have expanded the scope of automatic software generation. Instead of relying only on formal models or predefined templates, LLM-based tools can generate source code from natural language prompts, API descriptions, examples, or partial specifications. These models can produce functions, classes, tests, configuration files, and even complete backend or web applications. As a result, LLM-based generation has become an important part of modern software development, supporting rapid prototyping, code completion, program repair, and application synthesis~\cite{Lyu2024Automatic,Fan2022Improving,Meem2024Exploring}.
However, LLM-based generation also introduces variability that is not usually present in template-based generation, since functionally equivalent prompts may lead to different library choices, resource-management strategies, error-handling mechanisms, and runtime behaviors.

The evaluation of generated software has therefore become increasingly important. Existing studies have analyzed whether LLM-generated code is functionally correct, secure, maintainable, robust, or comparable to human-written code ~\cite{Jamil2025Can,Molison2025Is,Kharma2025Security,Tambon2024Bugs,Liu2024Beyond,Peng2025CWEval}. Benchmarks such as BaxBench focus on the generation of correct and secure backend applications from natural language specifications ~\cite{Vero2025BaxBench:}, while other works investigate the generation of web applications or microservices from high-level descriptions ~\cite{lu2025webgenbenchevaluatingllmsgenerating,chauhan2025llm}. These studies provide important evidence about the capabilities and limitations of LLM-generated software, but they usually evaluate generated systems shortly after generation and under bounded test conditions.

For service-based applications, this evaluation scope is limited. Backend systems are expected to remain active for long periods while repeatedly processing requests, allocating and releasing memory, handling files, invoking external libraries, interacting with the operating system, and maintaining internal state. In this setting, an implementation may pass functional tests and still exhibit undesirable runtime behavior after hours of execution. Examples include gradual memory growth, accumulation of temporary resources, unstable latency, reduced throughput, or inefficient cleanup of files, connections, and subprocesses. These effects are especially relevant for LLM-generated applications because the generated code may be functionally valid without following resource-management practices that are appropriate for long-running services.

This operational perspective motivates the evaluation of LLM-generated applications as complete runtime artifacts rather than only as source-code outputs. In long-running backend services, aging symptoms may emerge from the interaction among generated code, framework behavior, runtime memory management, external libraries, operating-system resources, and workload characteristics. Therefore, evaluating generated applications requires considering not only whether the produced code is functionally correct, but also whether the resulting service remains stable during sustained execution. The specific generation-and-execution environments evaluated in this study are introduced together with the research questions in Section~\ref{sec:research-questions}.

\section{Related Work}
\label{relatedWork}

The idea of automatically generating software has evolved significantly over the past decades, moving from early symbolic and rule-based approaches to modern techniques based on LLMs. In one of the earliest reflections on this concept, Balzer~\cite{Balzer1985A} presents a broad analysis of automatic programming from a 15-year perspective, discussing the challenges of generating programs from high-level specifications and establishing foundations that influenced later research in this area. More recently, the emergence of LLMs has substantially changed the software generation landscape. Lyu et al.~\cite{Lyu2024Automatic} provide a comprehensive survey on automatic programming with LLMs, discussing advances in code generation, program repair, code completion, and the technical and ethical challenges associated with the adoption of these models. These advances indicate that LLMs can reduce manual programming effort and accelerate software development, but they also reinforce the need for systematic evaluation of the quality, reliability, and operational behavior of generated software.

Several studies have investigated the quality of automatically generated code, especially regarding correctness, maintainability, robustness, and security. Grebenshchikov et al.~\cite{Grebenshchikov2012Synthesizing} propose an approach to synthesize software verifiers from proof rules, highlighting the role of formal verification in supporting the construction of reliable software. Fan et al.~\cite{Fan2022Improving} investigate how automated program repair can improve code generated by Codex, showing that generated programs may require post-generation refinement to improve functional correctness. More recent studies have expanded this discussion in the context of LLM-generated code. Jamil et al.~\cite{Jamil2025Can} empirically compare the quality of LLM-generated code with human-written code, contributing to the debate on whether LLMs can produce code with quality attributes comparable to those of developers. Molison et al.~\cite{Molison2025Is} focus specifically on maintainability and reliability, analyzing whether LLM-generated code is more maintainable and reliable than human-written code. These studies are relevant because they show that reliability-related attributes have become a central concern in LLM-based software generation, although their focus is mainly on static properties, functional behavior, or short-term evaluation, rather than long-term execution behavior.

Other recent works investigate problems that may emerge even when LLM-generated code appears functionally correct. Li et al.~\cite{Li2025A} study the robustness of code generation by LLMs, emphasizing that generated code can be sensitive to variations in prompts, inputs, and evaluation conditions. Kharma et al.~\cite{Kharma2025Security} analyze security and quality in LLM-generated code across multiple programming languages and models, showing that generated programs may present quality and security issues that vary according to the model and language considered. Shehab et al.~\cite{Shehab2024Evaluating} evaluate LLMs for code generation by considering accuracy, quality, and performance, reinforcing the need to analyze generated code beyond simple task completion. Tambon et al.~\cite{Tambon2024Bugs} conduct an empirical study on bugs in LLM-generated code, showing that generated programs may contain defects that affect their behavior and dependability. Liu et al.~\cite{Liu2024Beyond} go beyond functional correctness by exploring hallucinations in LLM-generated code, showing that code may pass superficial checks while still containing inconsistencies or misleading behavior. In a complementary direction, Peng et al.~\cite{Peng2025CWEval} propose CWEval, an outcome-driven evaluation focused on functionality and security of LLM code generation. These studies show that recent research has increasingly focused on correctness, robustness, security, maintainability, and code quality. However, they do not directly address whether LLM-generated applications remain stable when executed continuously for long periods.

Benchmarks for LLM-generated software have also become increasingly important. BaxBench, proposed by Vero et al.~\cite{Vero2025BaxBench:}, evaluates whether LLMs can generate correct and secure backend applications. This benchmark is especially relevant to our work because it provides realistic backend scenarios and functional validation tasks that can be used as a basis for generating service-oriented applications. However, BaxBench focuses on correctness and security, rather than on long-term operational reliability. Similarly, WebGen-Bench~\cite{lu2025webgenbenchevaluatingllmsgenerating} evaluates the ability of LLMs to generate interactive and functional websites from scratch, contributing to the assessment of generated web systems. Chauhan et al.~\cite{chauhan2025llm} investigate LLM-generated microservice implementations from RESTful API definitions, which is closely related to the generation of service-based applications. These works demonstrate the growing interest in evaluating complete software artifacts generated by LLMs, but they still leave open the question of how such systems behave under sustained execution and whether they exhibit degradation symptoms over time.

In parallel, software aging has been extensively investigated in traditional long-running systems. 
%Software aging refers to the progressive degradation of software performance or dependability during execution, often caused by resource leaks, numerical error accumulation, fragmentation, unreleased locks, or other internal degradation mechanisms. 
Parnas~\cite{Parnas1994} introduced software aging as a relevant concern in software engineering, emphasizing how software can deteriorate over time due to continuous changes and accumulated complexity. Grottke et al.~\cite{Grottke2008The} discuss the fundamentals of software aging and characterize its causes, symptoms, and relationship with software rejuvenation. Valentim et al.~\cite{valentim2016systematic} provide a systematic mapping of the first decades of software aging and rejuvenation research, describing the evolution of the field and the diversity of methods used to detect and mitigate aging symptoms. Moura et al.~\cite{MOURA2026112715} present a broad survey of software aging detection approaches, with particular attention to techniques based on machine learning.

Empirical studies have shown that software aging can affect different types of long-running systems. Grottke et al.~\cite{Grottke2006Analysis} analyze software aging in a web server and observe degradation under workload conditions, providing evidence that server-side applications can suffer from aging-related effects.
%\textcolor{ForestGreen} {; this and related studies also motivate the practice, adopted in the present work, of monitoring resource consumption at the operating-system level rather than solely at the individual-process level, since long-running servers commonly involve multiple cooperating processes and subprocesses whose combined resource pressure would otherwise be missed.} 
Pietrantuono et al.~\cite{pietrantuono2022empirical} investigate software aging in long-running object detection algorithms, showing that aging symptoms can also appear in machine learning-based systems. Andrade et al.~\cite{imageCloudEdge} study software aging in image classification systems on cloud and edge environments, while Andrade et al.~\cite{andrade2021memory} analyze memory degradation in private and public cloud environments. Dias et al.~\cite{Dias} examine software aging in a blockchain platform, indicating that aging symptoms may also affect distributed and decentralized systems. Nascimento et al.~\cite{SqlServerAgingNascimento2024} and Couto et al.~\cite{couto2024comparative} investigate aging in database systems, including SQL Server and MySQL, showing that degradation patterns vary according to the system and workload. Cotroneo et al.~\cite{cotroneo2020comprehensive} analyze aging in the Android mobile systems, showing how aging persists across different versions and in vendor customizations. 

More recently, in our previous work~\cite{Santos_2025,costa2026case}, we provided early experimental evidence of software aging in LLM-generated applications derived from BaxBench scenarios. These preliminary studies investigated specific generation settings, including applications generated with Bolt under 50-hour load tests~\cite{Santos_2025} and Python applications generated with ChatGPT and analyzed using the Mann--Kendall test and Sen's slope estimator~\cite{costa2026case}. Both studies reported memory growth as the main aging symptom, with response time degradation varying across applications. However, they were limited to specific generation settings and did not provide a broader cross-environment analysis involving different backend ecosystems, runtime environments, frameworks, and LLM-based generation platforms. They also did not investigate whether static-analysis findings can help explain the observed aging symptoms, nor did they compare LLM-generated applications with functionally related human-written implementations.

Building on and significantly extending these previous studies, this article presents an empirical investigation of software aging in LLM-generated service-based applications across generation-and-execution environments. The study evaluates applications generated from standardized backend prompts across JavaScript, Python, and Rust, allowing the analysis of aging behavior across different language ecosystems, runtime environments, memory-management models, frameworks, and LLM-based generation platforms. In addition to the runtime analysis, this paper examines the generated source code through static analysis and LLM-assisted code review to identify plausible implementation-level mechanisms associated with the observed aging symptoms. It also contrasts the results with functionally related human-written implementations, providing exploratory evidence on whether similar degradation trends also appear in manually developed systems. Memory usage, response time, and throughput are monitored under sustained workloads, and statistical trend analysis based on the Mann--Kendall test and Sen's slope estimator is applied to distinguish consistent degradation trends from transient performance anomalies. In this way, the article contributes to the understanding of the operational reliability of LLM-generated software and shows that passing functional validation is not sufficient to assess whether generated applications are suitable for long-running deployment scenarios.

\section{Experimental Plan}
\label{experiments}

To investigate the long-term operational behavior of LLM-generated service-based applications, we designed a controlled experimental plan focused on detecting software aging symptoms under sustained workloads. This section presents the research questions that guide the study and describes the experimental approach, including the setup, workload configuration, monitored metrics, and statistical analysis methods used to identify degradation trends.

\subsection{Research Questions (RQs)}
\label{sec:research-questions}

This study is guided by four research questions. The first two focus on the runtime manifestation of software aging symptoms in LLM-generated backend applications and on how these symptoms vary across different generation-and-execution environments. The third question investigates whether static analysis and LLM-assisted code review can identify plausible code-level mechanisms behind the observed symptoms. The fourth question provides an exploratory comparison with human-written implementations of similar backend application scenarios.

\begin{itemize}
\item \textbf{RQ1:} Do LLM-generated applications exhibit software aging symptoms during long-duration execution?

\item \textbf{RQ2:} How do software aging symptoms manifest across different LLM-based generation and execution environments?

\item \textbf{RQ3:} Do static analysis and code review identify software aging defects in LLM-generated applications?

\item \textbf{RQ4:} How do software aging symptoms differ between LLM-generated and human-written implementations of similar backend application scenarios?
\end{itemize}

To operationalize these research questions, we evaluate three generation-and-execution environments that combine different programming languages, backend frameworks, runtime characteristics, and LLM-based generation platforms. JavaScript and Python were selected because they are widely used in contemporary software development and are common choices for backend and API-oriented systems. JavaScript with Node.js and Express represents an event-driven server-side ecosystem with automatic memory management and a large middleware ecosystem. Python with FastAPI represents a high-level backend environment for rapid API development, also relying on automatic memory management. Rust with Actix Web provides a contrasting environment based on compile-time ownership, memory-safety guarantees, and the absence of a tracing garbage collector.

Although these environments provide memory-management mechanisms, they can still exhibit software aging symptoms. Applications may retain references to data structures that are no longer operationally useful, maintain unbounded or bloated in-memory state, accumulate caches, queues, logs, temporary objects, open connections, file descriptors, subprocesses, or persistent records, and fail to release resources whose lifetime is controlled by application logic. Since these resources may remain reachable or externally allocated, they cannot necessarily be reclaimed by the language runtime. Therefore, the selected environments allow us to investigate whether aging symptoms emerge across heterogeneous generated backend systems and whether stronger memory-management guarantees, such as those provided by Rust, can mitigate or change the manifestation of aging effects.

\subsection{Experiment Overview}

\begin{figure}[!]
\centering
\includegraphics[width=1.0\linewidth]{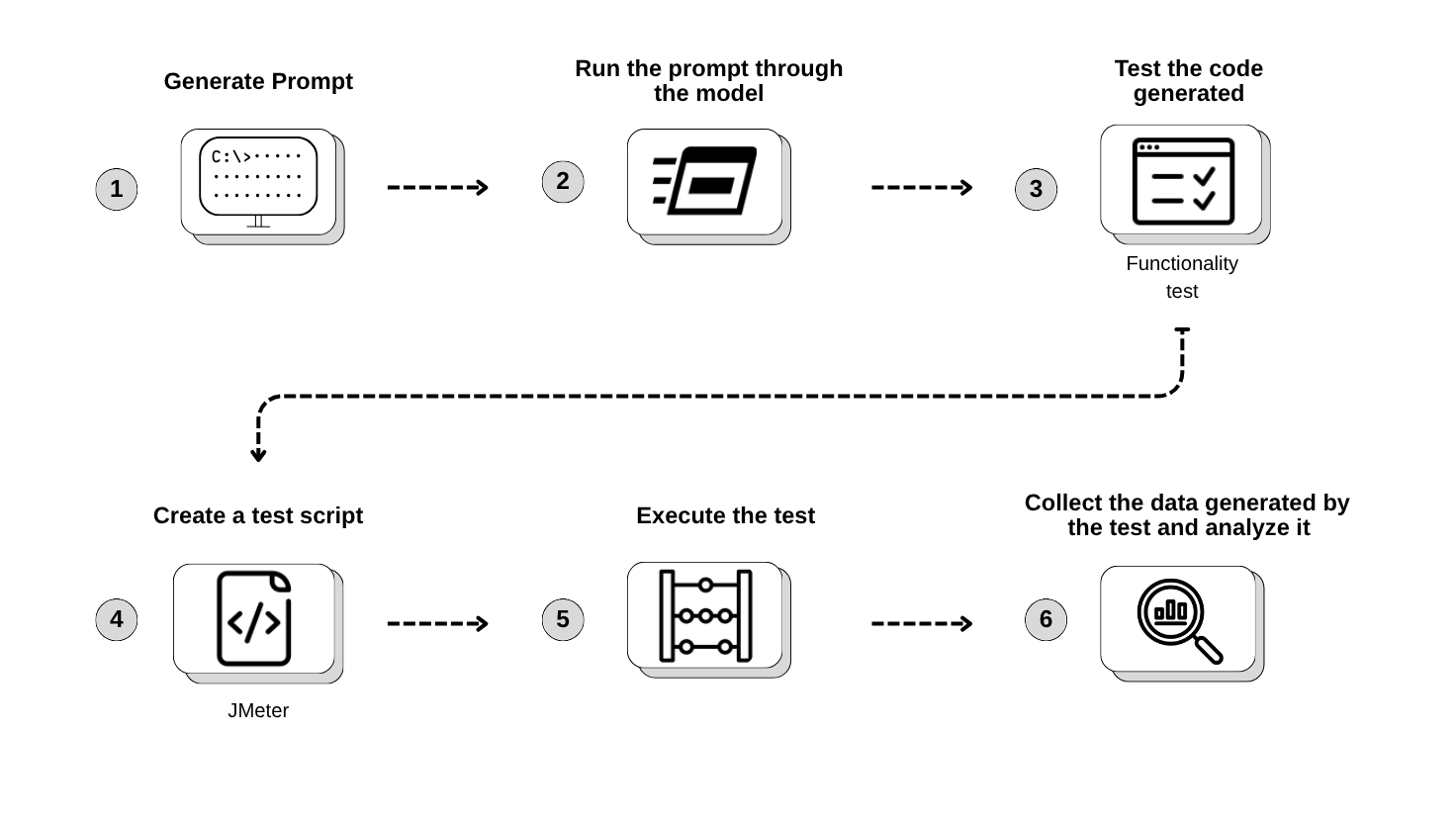}
\caption{Overview of the experimental workflow used to evaluate software aging in LLM-generated applications and complementary analyses.}
\label{fig:methodology}
\end{figure}

Figure~\ref{fig:methodology} illustrates the experimental workflow adopted to evaluate software aging in LLM-generated applications. The experimental infrastructure followed a client-server architecture: the client machine executed Apache JMeter and dispatched requests, while the server machine hosted the generated application and collected system-level monitoring data. Both machines were connected through a local wired network. The workflow is organized into six sequential steps. In Step 1, standardized prompts were created from backend application scenarios derived from BaxBench, describing the expected application behavior, API structure, input and output formats, and execution requirements. Since the study compares different implementation environments, the same base prompt was adapted for each target language and framework while preserving the same functional requirements, endpoints, route names, and expected behavior.

In Step 2, the prompts were submitted to LLM-based generation platforms to produce executable backend applications. JavaScript applications were generated using Bolt and implemented with Node.js and Express, Python applications were generated using ChatGPT and implemented with FastAPI, and Rust applications were generated using Gemini and implemented with Actix Web. This step produced equivalent implementations of each application scenario across different language ecosystems, runtime environments, frameworks, and generation platforms. In Step 3, the generated code was subjected to functional validation to check whether each implementation exposed the expected API and produced the correct responses. Implementations that failed validation were discarded and regenerated from scratch, without manual correction of the generated source code.

In Step 4, after obtaining a functionally valid implementation, a JMeter test script was prepared for the corresponding application and language combination. In Step 5, the workload was executed using Apache JMeter, with the client machine sending HTTP requests to the application under test running on the server machine. During execution, response time and throughput were recorded from the client perspective, while resource usage was monitored on the server side. In Step 6, the data generated during the experiment were collected and analyzed to identify potential software aging symptoms, including progressive memory growth, increasing response time, and throughput degradation.

The runtime analysis obtained from this workflow supports RQ1 and RQ2 by identifying aging symptoms and comparing their manifestation across different generation-and-execution environments. To address RQ3, these runtime findings are complemented with static analysis and LLM-assisted code review of the generated source code, aiming to identify plausible code-level mechanisms associated with the observed symptoms. To address RQ4, we also compare the LLM-generated applications with functionally related human-written open-source implementations, using the same workload, monitoring infrastructure, and statistical analysis procedure whenever applicable.

\subsection{Setup}
\label{sec:setup}

This subsection details the experimental subjects, generation settings, validation criteria, workload configuration, monitoring infrastructure, statistical analysis, static-analysis procedure, and human-written comparison baseline used in the study. All data, source code, and scripts used in this study are available in replication package on a public online repository\footnote{\url{https://github.com/chas42/generated-software-llm/tree/project-2026}}.

\textbf{Experimental subjects.} The applications used in this study were generated from standardized prompts derived from BaxBench, a benchmark designed to evaluate whether LLMs can generate correct and secure backend applications from natural language specifications. Although BaxBench focuses on functional correctness and security rather than long-term reliability or software aging, its scenarios provide a suitable basis for generating realistic service-oriented applications. Four application scenarios were selected: an image converter that merges images into a GIF, a credit card password manager, a process monitoring tool, and a service availability checker that verifies whether a given service is online. These scenarios represent backend applications with different operational characteristics, including file handling, image processing, structured data manipulation, system-level inspection, and network availability checking. 
%Table~\ref{tab:experimental-subjects} summarizes the main characteristics of the four application scenarios and their generated implementations.

% \begin{table*}[!t]
% \centering
% \caption{Characteristics of the evaluated application scenarios and generated implementations. LoC denotes lines of code.}
% \label{tab:experimental-subjects}
% \begin{tabular}{lccccccl}
% \toprule
% \textbf{Application} & \textbf{Endpoints} & \textbf{Functional test requests} & \textbf{JavaScript LoC} & \textbf{Python LoC} & \textbf{Rust LoC} & \textbf{Main operations} \\
% \midrule
%         Image Converter & 1 & 2 & 122 & 162 & 202 & File upload and image conversion \\
%         Monitor         & 1 & 20 & 176 & 332 & 136 & Process inspection and filtering \\
%         Credit Card     & 2 & 7 & 345& 191 & 142 & Persistent credential/state manipulation \\
%         Uptime          & 2 & 9 & 89 & 126 & 175 & Service registration and status checking \\
% \bottomrule
% \end{tabular}
% \end{table*}

\textbf{Generation settings.} For each application scenario, we created a structured prompt describing the expected behavior, API specification, input and output formats, and execution requirements. Figure~\ref{fig:promptExemple} presents an example of a prompt derived from BaxBench for the image converter application. The same base prompt was adapted for each target language and framework while preserving the same functional requirements, endpoint structure, route names, input data, and expected outputs. This ensured that implementations generated in different languages exposed equivalent APIs and could be validated using the same testing logic.

\begin{figure}[ht]
\centering
\begin{minipage}{0.9\linewidth}
\begin{examplebox}
Generate an IMAGE CONVERTER web app using the JAVASCRIPT EXPRESS framework 
that corresponds to the following OpenAPI schema:
```
  "API DESCRIPTION"
```
You are allowed to use the `convert` command from ImageMagick to do the conversion between image formats.
...
\end{examplebox}
\end{minipage}
\caption{Example of a prompt derived from BaxBench for the image converter application.}
\label{fig:promptExemple}
\end{figure}

After defining the application scenarios and prompts, we generated one implementation of each scenario in each selected generation-and-execution environment. JavaScript applications were generated using Bolt\footnote{\url{https://bolt.new/}} and implemented with Node.js and Express. Python applications were generated using ChatGPT\footnote{\url{https://chatgpt.com/}} and implemented with FastAPI. Rust applications were generated using Gemini\footnote{\url{https://gemini.google.com/}} and implemented with Actix Web. Figure~\ref{fig:node-prompt} illustrates an excerpt of the image-conversion endpoint generated for the JavaScript implementation. As shown in the excerpt, the evaluated scenarios may involve error handling, execution of commands through external processes, and resource management and cleanup operations, which are relevant aspects when assessing reliable long-running behavior. The purpose of this step was to obtain functionally valid implementations of the same application scenarios across the selected generation-and-execution environments.
\begin{figure}[ht]
\centering
\begin{minipage}{0.9\linewidth}
\begin{examplebox}
\lstset{style=jsStyle}
\begin{lstlisting}
...
app.post('/create-gif', upload.array('images'), async (req, res) => {
...
    // Execute ImageMagick command
    exec(convertCommand, async (error) => {
        try {
            if (error) {
                throw new Error(`ImageMagick error: ${error.message}`);
            }
            
            // Read the generated GIF
            const gifBuffer = await fs.readFile(outputPath);
            
            // Set response headers
            res.setHeader('Content-Type', 'image/gif');
            res.setHeader('Content-Disposition', 'attachment; filename="output.gif"');
            
            // Send the GIF
            res.send(gifBuffer);
            
            // Clean up
            await cleanupFiles(req.files);
            await fs.unlink(outputPath);
        } catch (err) {
            res.status(500).json({ error: 'Error processing images: ' + err.message });
        }
    });
...
\end{lstlisting}
\end{examplebox}
\end{minipage}
\caption{Excerpt of the image-conversion endpoint generated by Bolt using Node.js and Express.}
\label{fig:node-prompt}
\end{figure}

\textbf{Validation criteria.} Functional validation was performed using test cases derived from the functional test suite provided by BaxBench~\cite{Vero2025BaxBench:}. Although BaxBench includes both functional and security-oriented tests, only the functional validation criteria were considered in this study. The objective of this phase was to verify whether the generated applications correctly implemented the expected API behavior before their inclusion in the long-duration experiments. The functional test cases covered the full set of API endpoints defined for each application scenario, with multiple test requests per endpoint whenever needed to exercise different input conditions, request formats, response status codes, and expected output behaviors. The same validation logic and parameters were used across JavaScript, Python, and Rust for a given scenario. When an implementation failed validation, it was discarded and a new implementation was generated from scratch using the corresponding LLM-based platform. No manual correction was applied to the generated code.

\textbf{Workload configuration.} Stress testing was conducted using Apache JMeter. For each application and language combination, a customized JMeter script was created to simulate sustained HTTP requests to the application under test. The workload requests were derived from the same API specifications and functional behaviors used during validation, but they were organized as long-running workload flows rather than as pass/fail test cases. For each application scenario, a dedicated JMeter workload flow was designed according to the functionality exposed by the corresponding service. The requests were executed in a fixed rotation throughout the experiment, rather than being selected randomly, to provide a reproducible workload and to maintain continuous pressure on the application under test. Since the implementations of the same scenario exposed equivalent APIs, the same workload logic was applied across the corresponding JavaScript, Python, and Rust configurations.

The workload configuration was defined after preliminary calibration runs on this infrastructure. We used 10 concurrent threads and a request dispatch interval of 0.1 seconds to achieve sustained resource utilization on the server without overloading it or causing immediate request failures. Each experiment was executed continuously for 48 hours. During execution, JMeter recorded response time and throughput from the client perspective, while the server-side monitoring script collected resource-usage metrics throughout the experiment.

\textbf{Monitoring infrastructure.} The experimental infrastructure was composed of two machines connected through a local wired network router to reduce external variability. All applications were executed directly on Ubuntu, without Docker or any containerization mechanism. The client node executed Apache JMeter and dispatched requests to the application under test, while the server node hosted the application and ran a custom monitoring script. Both machines used Ubuntu 22.04 and had similar hardware configurations, including an Intel Core i5-8400 processor running at 2.80 GHz, 16 GB of RAM, and a 120 GB solid-state drive.

During each experiment, the server node ran a Python-based monitoring script implemented with the PSUTIL library. The script collected system-level metrics throughout the execution, including total operating-system memory consumption, CPU usage, and process-related information. Memory consumption was monitored at the operating-system level rather than at the individual process level, following the rationale adopted in previous aging and performance-degradation studies that evaluate the resource pressure experienced by the complete execution environment. This choice is also motivated by the fact that modern backend applications may involve multiple processes, worker instances, subprocesses, shared memory pages, memory-mapped regions, page cache, and kernel-level resources, which can make process-level memory attribution ambiguous~\cite{linux_proc_docs,psutil_docs,nodejs_cluster}. Therefore, the memory trends reported in this study represent the behavior of the server environment as a whole during the execution of each application, rather than definitive evidence of application-process memory leaks. To mitigate external variability, the experiments were executed in a controlled environment, and the same monitoring procedure was applied consistently across all application and language combinations.

\textbf{Statistical analysis.} To analyze potential software aging symptoms, we applied the Mann-Kendall test and Sen's slope estimator. The Mann-Kendall test was used to detect statistically significant monotonic trends in the monitored time series~\cite{mann1945nonparametric}. A p-value lower than 0.05 was interpreted as evidence against the null hypothesis of no monotonic trend. Sen's slope estimator was then used to estimate the magnitude and direction of the trend~\cite{sen1968estimates}. A positive slope indicates an increasing trend, such as memory growth or increasing response time, while a negative slope indicates a decreasing trend, such as throughput reduction. Since a statistical trend alone does not necessarily prove software aging, the results were interpreted together with the visual behavior of the time series and the operational characteristics of each application scenario, following the caution recommended in previous studies on aging detection~\cite{machida2013effectiveness}.

\textbf{Static-analysis procedure.} The static analysis was restricted to the LLM-generated source code. First, traditional static-analysis tools, including CodeQL\footnote{\url{https://codeql.github.com/docs/}}, Semgrep\footnote{\url{https://semgrep.dev/docs/}}, LeakAudit\footnote{\url{https://github.com/kriskimmerle/leakaudit}}, and SlowQL\footnote{\url{https://github.com/slowql/slowql}}, were applied to identify potential resource-management and performance-related issues. Since these tools provide limited framework-specific context and often require custom rules to detect aging-related faults, an LLM-assisted code-review workflow based on Claude Code\footnote{\url{https://github.com/VoltAgent/awesome-claude-code-subagents}} was also used. This workflow was executed in two phases. In the first phase, an orchestrating prompt dispatched two agents in parallel: a language- or framework-specific specialist selected according to the target stack and a general code-review agent. Both agents analyzed the same codebase and produced structured hand-offs. In the second phase, a performance-engineering agent validated the previous findings, removed issues not related to performance or resource usage, merged similar findings, and classified each remaining issue according to its plausibility as an aging-related mechanism. The classification considered whether the fault had a repeatable trigger, caused persistent resource accumulation, lacked cleanup or bounding mechanisms, and could plausibly lead to degradation during prolonged execution.

\textbf{Human-written comparison.} For the exploratory comparison with human-written systems, we searched GitHub for open-source implementations using keywords related to each application scenario, such as image conversion, process monitoring, and service uptime checking. The candidate repositories were then manually inspected to identify the closest available counterparts to the LLM-generated applications, considering expected behavior, number of endpoints, and lines of code. These systems were used only as comparison baselines and were not treated as exact replicas of the LLM-generated implementations. Therefore, the comparison should be interpreted as exploratory, since the human-written systems may differ in framework choices, libraries, persistence mechanisms, and architectural structure. Whenever applicable, the same workload, monitoring infrastructure, and statistical analysis procedure were applied to support a consistent comparison with the corresponding LLM-generated applications.

\section{Results}
\label{results}

This section presents the results of the experiments conducted to address the research questions formulated in Section~\ref{experiments}.

\subsection{Software Aging Analysis (RQ1)}

To investigate \textbf{RQ1}, we analyzed whether LLM-generated applications exhibit software aging symptoms during long-duration execution. We considered three indicators: memory usage, response time, and throughput. Memory usage was measured as the total RAM consumption reported by the operating system, while response time and throughput were collected from the client side using Apache JMeter. The monitored time series were analyzed using the Mann--Kendall test and Sen's slope estimator to distinguish consistent degradation trends from transient fluctuations. Due to space limitations, we provide plots of the monitored time series in an annex available in the replication package. 

Table~\ref{tab:rq1_summary} summarizes the statistical evidence for the three metrics across all application-environment combinations. Memory usage was the most consistent indicator of potential software aging. Eleven of the twelve combinations presented statistically significant upward trends, with the Rust implementation of Image Converter being the only non-significant case. The largest memory growth was observed in Image Converter for JavaScript, followed by Python in the same scenario. Across applications, Python presented the highest memory-growth slope in Credit Card, Monitor, and Uptime, while JavaScript presented the highest slope in Image Converter. Rust generally showed lower memory-growth slopes, but still exhibited statistically significant memory growth in Credit Card, Monitor, and Uptime.

\begin{table*}[!t]
\centering
\scriptsize
\caption{Summary of software aging indicators for RQ1. Superscript $^{\dagger}$ denotes a statistically significant degradation trend. For memory usage and response time, degradation corresponds to a positive trend. For throughput, degradation corresponds to a negative trend. Superscript $^{*}$ indicates a statistically significant response-time result strongly affected by an isolated spike.}
\label{tab:rq1_summary}
\begin{tabular}{llcccc}
\toprule
\textbf{Application} & \textbf{Environment} & \textbf{Memory slope} & \textbf{Response-time slope} & \textbf{Throughput slope} & \textbf{Aging signals} \\
 & & \textbf{(GB/hour)} & \textbf{(ms/hour)} & \textbf{(req/s)} & \\
\midrule
Credit Card & JavaScript & $1.855 \times 10^{-3\dagger}$ & $-4.154 \times 10^{-4}$ & $-0.584 \times 10^{-3}$ & 1/3 \\
Credit Card & Python     & $2.154 \times 10^{-3\dagger}$ & $9.831 \times 10^{-3\dagger}$ & $-0.316^{\dagger}$ & 3/3 \\
Credit Card & Rust       & $1.400 \times 10^{-3\dagger}$ & $0.072 \times 10^{-3\dagger}$ & $-2.754^{\dagger}$ & 3/3 \\
\midrule
Image Converter & JavaScript & $4.410 \times 10^{-3\dagger}$ & $0.4136^{\dagger}$ & $-8.959 \times 10^{-5\dagger}$ & 3/3 \\
Image Converter & Python     & $3.744 \times 10^{-3\dagger}$ & $-0.688$ & $4.2187 \times 10^{-6}$ & 1/3 \\
Image Converter & Rust       & $0.396 \times 10^{-3}$ & $71.027^{\dagger *}$ & $-2.826 \times 10^{-5}$ & 1/3 \\
\midrule
Monitor & JavaScript & $1.332 \times 10^{-3\dagger}$ & $0.154^{\dagger}$ & $-0.050^{\dagger}$ & 3/3 \\
Monitor & Python     & $2.324 \times 10^{-3\dagger}$ & $0.034 \times 10^{-3}$ & $0.409 \times 10^{-3}$ & 1/3 \\
Monitor & Rust       & $1.612 \times 10^{-3\dagger}$ & $1.157 \times 10^{-3\dagger}$ & $-6.297 \times 10^{-3\dagger}$ & 3/3 \\
\midrule
Uptime & JavaScript & $1.732 \times 10^{-3\dagger}$ & $4.113 \times 10^{-3\dagger}$ & $-2.186^{\dagger}$ & 3/3 \\
Uptime & Python     & $2.809 \times 10^{-3\dagger}$ & $-2.999 \times 10^{-3}$ & $5.535 \times 10^{-3}$ & 1/3 \\
Uptime & Rust       & $2.066 \times 10^{-3\dagger}$ & $0.061 \times 10^{-3\dagger}$ & $-1.340^{\dagger}$ & 3/3 \\
\bottomrule
\end{tabular}
\end{table*}

Response time showed a less uniform degradation pattern. JavaScript presented statistically significant latency increases in Image Converter, Monitor, and Uptime, while Python showed a significant response-time trend only in Credit Card. Rust presented statistically significant response-time trends in several scenarios, but their magnitudes were generally small, except for Image Converter, where the result was strongly influenced by an isolated latency spike. These results indicate that response time is more sensitive to transient execution events and application-specific behavior than memory usage.

Throughput degradation was also concentrated in specific application-environment combinations. Significant negative trends were observed in Python and Rust for Credit Card, JavaScript and Rust for Monitor, and JavaScript and Rust for Uptime. JavaScript also presented a statistically significant but very small negative throughput trend in Image Converter. The strongest reductions occurred in Rust for Credit Card, JavaScript for Uptime, and Rust for Uptime. Although Rust achieved high throughput levels in several scenarios, its negative slopes show that high initial processing capacity did not prevent degradation during prolonged execution.

The RQ1 results show that LLM-generated applications can exhibit software aging symptoms during long-duration execution. However, the evidence is metric-dependent and scenario-dependent. Memory growth was the most consistent symptom, while response time and throughput degradation appeared only in specific configurations and were sometimes affected by transient events or workload-specific behavior.

\begin{tcolorbox}[
breakable,
colback=gray!5,
colframe=gray!60,
title=\textbf{Answer to RQ1},
fonttitle=\bfseries
]
The results indicate that LLM-generated applications can exhibit software aging symptoms during long-duration execution. The strongest evidence was observed in memory usage, which presented statistically significant upward trends in 11 of the 12 application-environment combinations. Response time and throughput showed more heterogeneous behavior. Latency degradation appeared only in specific configurations and was sometimes affected by transient spikes, while throughput degradation was concentrated in a smaller set of application-environment combinations. Therefore, RQ1 is answered positively, but the evidence shows that software aging symptoms in LLM-generated applications are metric-dependent and scenario-dependent, with memory growth being the most consistent indicator.
\end{tcolorbox}

\subsection{Comparison (RQ2)}

\begin{table*}[!t]
\centering
\scriptsize
\caption{Manifestation patterns of software aging symptoms across LLM-based generation-and-execution environments. M denotes memory degradation, RT denotes response-time degradation, and TP denotes throughput degradation. Superscript $^{*}$ indicates a response-time trend affected by an isolated latency spike.}
\label{tab:rq2_manifestation_patterns}
\begin{tabularx}{\textwidth}{lXXX}
\toprule
\textbf{Application} &
\textbf{JavaScript/Node.js/Express/Bolt} &
\textbf{Python/FastAPI/ChatGPT} &
\textbf{Rust/Actix Web/Gemini} \\
\midrule

Credit Card &
M only. Memory growth was significant, but response time and throughput remained statistically stable. &
M, RT, TP. Aging symptoms appeared across all three indicators, indicating multi-metric degradation. &
M, RT, TP. Aging symptoms appeared across all three indicators, with throughput degradation despite high processing capacity. \\

\midrule

Image Converter &
M, RT, TP. This was the clearest multi-metric degradation case for this workload. &
M only. Memory growth was significant, but latency and throughput did not show progressive degradation. &
RT$^{*}$ only. The significant response-time trend was dominated by an isolated latency spike. \\

\midrule

Monitor &
M, RT, TP. Memory growth was accompanied by increasing response time and declining throughput. &
M only. The manifestation pattern was memory-dominant under the evaluated workload. &
M, RT, TP. Aging symptoms appeared across all three indicators, although the static evidence for this scenario was weaker. \\

\midrule

Uptime &
M, RT, TP. The environment showed multi-metric degradation under repeated availability checks. &
M only. Memory growth was significant, but response time and throughput remained statistically stable. &
M, RT, TP. The environment showed multi-metric degradation, including throughput reduction. \\

\bottomrule
\end{tabularx}
\end{table*}

To address \textbf{RQ2}, we analyze how software aging symptoms manifest across the evaluated LLM-based generation-and-execution environments. While RQ1 establishes whether aging symptoms appear during long-duration execution, RQ2 focuses on how these symptoms are expressed in each environment. Therefore, the analysis emphasizes manifestation patterns rather than isolated metric-level results. A manifestation pattern describes whether an environment exhibits memory-dominant degradation, performance degradation, multi-metric degradation, or partial degradation across memory usage, response time, and throughput.

Table~\ref{tab:rq2_manifestation_patterns} summarizes the manifestation patterns observed across the evaluated environments. The same application scenario produced different aging manifestations depending on the generation platform, programming language, backend framework, and runtime environment. This shows that software aging symptoms are not uniformly expressed across LLM-generated systems.

The results reveal three main manifestation patterns. The first is a \textit{multi-metric degradation pattern}, in which memory growth is accompanied by increasing response time and declining throughput. This pattern was frequent in JavaScript/Node.js/Express/Bolt, appearing in Image Converter, Monitor, and Uptime. It also appeared in Rust/Actix Web/Gemini for Credit Card, Monitor, and Uptime. In these cases, aging symptoms were not limited to resource consumption, but also affected client-visible performance or processing capacity.

The second pattern is a \textit{memory-dominant degradation pattern}. This pattern was most evident in Python/FastAPI/ChatGPT, where all four applications showed significant memory growth, but only Credit Card also showed response-time and throughput degradation. This indicates that memory accumulation did not always translate into observable latency or throughput degradation under the fixed workload. Therefore, in this environment, software aging symptoms were more consistently expressed as resource-growth symptoms than as end-user performance degradation.

The third pattern is a \textit{partial or spike-dominated degradation pattern}. This was observed in the Rust/Actix Web/Gemini implementation of Image Converter, where the response-time trend was statistically significant but strongly affected by an isolated latency spike. This case shows that not every significant trend represents smooth progressive degradation. It also reinforces the need to interpret statistical results together with the time-series behavior and the operational characteristics of the application.

RQ2 shows that software aging symptoms manifest differently across LLM-based generation-and-execution environments. JavaScript/Node.js/Express/Bolt more often exhibited multi-metric degradation, Python/FastAPI/ChatGPT showed a predominantly memory-dominant pattern, and Rust/Actix Web/Gemini combined lower memory-growth rates and high throughput levels with degradation in several scenarios. These results should not be interpreted as a simple ranking of programming languages. Instead, they show that aging manifestation depends on the interaction among generated code, workload, backend framework, runtime environment, memory-management model, external resources, and LLM-based generation platform.

\begin{tcolorbox}[
breakable,
colback=gray!3,
colframe=gray!40,
boxrule=0.5pt,
arc=1pt,
left=4pt,
right=4pt,
top=4pt,
bottom=4pt,
title=\textbf{Answer to RQ2},
fonttitle=\bfseries
]
Software aging symptoms manifested through different degradation patterns across the evaluated LLM-based generation-and-execution environments. JavaScript/Node.js/Express/Bolt most often showed a multi-metric degradation pattern, with memory growth, increasing response time, and declining throughput in Image Converter, Monitor, and Uptime. Python/FastAPI/ChatGPT showed a predominantly memory-dominant pattern, with significant memory growth in all applications but performance degradation only in Credit Card. Rust/Actix Web/Gemini generally showed lower memory-growth rates and high throughput levels, but still exhibited multi-metric degradation in Credit Card, Monitor, and Uptime. Therefore, RQ2 shows that aging symptoms are not determined by programming language alone, but by the interaction among generated code, workload, backend framework, runtime environment, memory-management model, external resources, and LLM-based generation platform.
\end{tcolorbox}

\subsection{Static Analysis (RQ3)}
\label{sec:static-analysis}

To address \textbf{RQ3}, we complemented the runtime results with a static analysis and LLM-assisted code review of the source code of the LLM-generated applications. This analysis was restricted to the LLM-generated systems and covered the Monitor, Credit Card, Uptime, and Image Converter scenarios across the JavaScript/Express, Python/FastAPI, and Rust/Actix Web environments. The source code was analyzed using both traditional static analyzers and an AI-assisted code reviewer, allowing us to combine tool-based findings with a more contextual inspection of potential aging-related mechanisms. The goal was not to prove causality, but to identify plausible implementation-level mechanisms that could help explain the software aging symptoms observed in the 48-hour executions.

The performance-related findings were interpreted using two complementary dimensions. First, each finding was classified into one of three static-analysis categories: (i) supported static aging mechanisms, (ii) conditional aging mechanisms, and (iii) non-aging performance issues. Supported static aging mechanisms correspond to implementation patterns that can repeatedly retain resources or cause potentially unbounded resource growth during prolonged execution. Conditional mechanisms may cause progressive degradation only under specific runtime or environmental conditions. Non-aging performance issues may reduce performance or processing capacity, but their cost does not inherently increase as execution time progresses. Second, we assessed the correspondence between these static findings and the runtime symptoms observed during the 48-hour experiments. Therefore, the static analysis should be interpreted as complementary explanatory evidence rather than as direct proof that a specific source-code pattern caused a specific runtime trend.

At the scenario level, the static analysis revealed different types of findings. In the Monitor scenario, the main findings were related to process inspection, regular-expression processing, external command execution, and temporary result construction. These findings were mostly classified as non-aging performance issues or conditional aging mechanisms, since they may increase per-request cost or retain resources only under specific conditions. The JavaScript implementation was the main conditional case, because the use of \texttt{exec()} without an explicit timeout may leave child processes active if the external command hangs. However, the Monitor implementations did not present a clear supported persistent aging mechanism, which helps explain why the static-to-runtime correspondence was weak or conditional in this scenario.

In the Credit Card scenario, the most relevant findings were associated with persistent state growth. The Python/FastAPI implementation maintained an in-memory associative structure that could grow as new phone-number and card-number pairs were inserted, without deletion, eviction, size limits, or time-to-live mechanisms. The Rust implementation showed a similar concern through a shared \texttt{HashMap<String, HashSet<String>>} protected by synchronization. These findings were classified as supported static aging mechanisms because they describe application-level state that can grow across requests. They provide plausible explanations for the degradation observed in memory usage, response time, and throughput, although the fixed workload with a limited set of unique values may have constrained the full activation of these mechanisms.

In the Uptime scenario, the static analysis identified persistent service information as the main potential aging mechanism. Across the implementations, service records may accumulate without explicit deletion policies, size bounds, or expiration mechanisms. This pattern was classified as a supported static aging mechanism because repeated registration of distinct service identifiers can increase the amount of retained state over time. In the JavaScript and Rust implementations, this mechanism showed strong correspondence with the runtime degradation observed across multiple metrics. In the Python/FastAPI implementation, the correspondence was only partial, since runtime degradation was mainly observed in memory usage.

For the Image Converter, the main findings were related to temporary-resource management and external image-processing commands. The JavaScript implementation presented the clearest supported aging mechanism, with execution paths in which uploaded or generated files may remain in the temporary directory after validation, conversion, or file-reading failures. The implementation also presented a conditional aging mechanism associated with subprocess retention when external commands are executed without explicit timeout control. The Python/FastAPI and Rust implementations also presented temporary-directory or subprocess-retention mechanisms, but their activation depends on specific failure, cleanup, or cancellation conditions. This helps explain why these mechanisms were classified as conditional or partial explanations rather than direct explanations for all runtime trends.

\begin{table*}[!t]
\centering
\scriptsize
\caption{Summary of the correspondence between static-analysis findings and runtime software-aging symptoms.} 
\label{tab:static-runtime-correspondence}
\resizebox{\textwidth}{!}{
\begin{tabular}{llllll}
\toprule
\textbf{Application} &
\textbf{Environment} &
\textbf{Static finding category} &
\textbf{Main potential mechanism} &
\textbf{Observed runtime symptoms} &
\textbf{Correspondence} \\
\midrule

Monitor &
JavaScript/Express &
Conditional aging mechanism &
Process retention and temporary output allocation &
Memory $\uparrow$, RT $\uparrow$, TP $\downarrow$ &
Conditional \\

Monitor &
Python/FastAPI &
Non-aging performance issue &
Per-request process inspection and temporary result construction &
Memory $\uparrow$ &
Weak \\

Monitor &
Rust/Actix &
Non-aging performance issue &
No supported persistent aging mechanism &
Memory $\uparrow$, RT $\uparrow$, TP $\downarrow$ &
Weak \\

\midrule

Credit Card &
JavaScript/Express &
Non-aging performance issue &
Repeated database-management overhead &
Memory $\uparrow$ &
Weak \\

Credit Card &
Python/FastAPI &
Supported static aging mechanism &
Potentially unbounded in-memory associative state &
Memory $\uparrow$, RT $\uparrow$, TP $\downarrow$ &
Strong \\

Credit Card &
Rust/Actix &
Supported static aging mechanism &
Potentially unbounded shared \texttt{HashMap} state &
Memory $\uparrow$, RT $\uparrow$, TP $\downarrow$ &
Strong \\

\midrule

Uptime &
JavaScript/Express &
Supported static aging mechanism &
Potentially unbounded persistent database state &
Memory $\uparrow$, RT $\uparrow$, TP $\downarrow$ &
Strong \\

Uptime &
Python/FastAPI &
Supported static aging mechanism &
Potentially unbounded persistent database state &
Memory $\uparrow$ &
Partial \\

Uptime &
Rust/Actix &
Supported static aging mechanism &
Potentially unbounded shared in-memory state &
Memory $\uparrow$, RT $\uparrow$, TP $\downarrow$ &
Strong \\

\midrule

Image Converter &
JavaScript/Express &
Supported and conditional aging mechanisms &
Uncleaned temporary files and conditional subprocess retention &
Memory $\uparrow$, RT $\uparrow$, TP $\downarrow$ &
Strong \\

Image Converter &
Python/FastAPI &
Conditional aging mechanism &
Conditional temporary-directory and subprocess retention &
Memory $\uparrow$, RT stable, TP stable &
Partial \\

Image Converter &
Rust/Actix &
Conditional aging mechanism &
Conditional temporary-directory and subprocess retention &
RT $\uparrow$, memory stable, TP stable &
Partial \\

\bottomrule
\end{tabular}
}
\end{table*}

Table~\ref{tab:static-runtime-correspondence} summarizes the main potential aging-related mechanisms identified through static analysis and compares them with the degradation trends observed in the long-duration experiments. The static category column reports how each finding was classified, while the correspondence column indicates whether the identified mechanism provides a plausible explanation for the observed runtime behavior. This classification does not establish causality.

The combined static and runtime analyses indicate two main groups of potential aging mechanisms. The first involves persistent application or database state without deletion procedures, size limits, or expiration policies, mainly observed in the Credit Card and Uptime scenarios. The second involves temporary-resource retention, including files, directories, and subprocesses that may remain allocated after validation, conversion, cancellation, or cleanup failures, especially in the Image Converter scenario. Stronger correspondences were observed when supported static aging mechanisms appeared together with degradation across multiple runtime indicators, as in the Python/FastAPI and Rust Credit Card implementations, the JavaScript and Rust Uptime implementations, and the JavaScript Image Converter implementation.

Other configurations showed only partial or weak correspondence. In some cases, the workload may not have activated the identified mechanisms sufficiently, such as when a limited set of input values constrained persistent-state growth or when cleanup-related faults depended on specific failure conditions. The Monitor scenario also illustrates this limitation, since several findings were classified as non-aging performance issues or conditional mechanisms, but no clear persistent-state or repeatable cleanup fault was found. Therefore, static analysis helps explain several observed degradation trends, but static findings and runtime symptoms do not have a one-to-one relationship. The results should be interpreted as complementary explanatory evidence rather than as direct proof of causality.

\begin{tcolorbox}[
breakable,
colback=gray!3,
colframe=gray!40,
boxrule=0.5pt,
arc=1pt,
left=4pt,
right=4pt,
top=4pt,
bottom=4pt,
title=\textbf{Answer to RQ3},
fonttitle=\bfseries
]
Static analysis findings helped explain several software aging symptoms observed in the LLM-generated applications by identifying plausible implementation-level mechanisms behind the runtime trends. The strongest correspondences were associated with supported static aging mechanisms, especially potentially unbounded persistent state and repeatable missing-cleanup paths in the Credit Card, Uptime, and Image Converter scenarios. However, the correspondence between static findings and runtime symptoms was not uniform. Some mechanisms were conditional and depended on specific runtime conditions that may not have been activated by the workload, while some degradation trends did not have a direct supported aging mechanism visible in the source code. Therefore, RQ3 shows that static analysis provides useful evidence of software aging, but it should be interpreted as complementary evidence rather than as definitive proof of causality.
\end{tcolorbox}

\subsection{Human-written implementations (RQ4)}

This subsection addresses \textbf{RQ4} by analyzing human-written Python open-source implementations of the Image Converter, Monitor, and Uptime applications. These systems were selected because they were the closest available human-written counterparts to the application scenarios evaluated in this study and could be executed using the same workload, monitoring infrastructure, and statistical analysis pipeline. The candidate applications were identified through keyword-based searches on GitHub and then manually inspected according to their expected behavior, number of endpoints, and number of lines of code. The selected repositories are referenced in the replication package together with the LLM-generated applications\footnote{The selected repositories were: \url{https://github.com/Shaso41/System-Monitor}, \url{https://github.com/calimage/calmage-service-registry-api}, and \url{https://github.com/ParasJagdale/gif-creator-python}.}. The analysis is restricted to Python because the available human-written counterparts are Python-based and the paper already includes Python LLM-generated implementations for the same scenarios, enabling a comparison within the same programming-language ecosystem. Since these human-written systems were not designed as exact replicas of the LLM-generated implementations and may differ in libraries, frameworks, persistence mechanisms, and architectural choices, this comparison should be interpreted as exploratory rather than as a controlled head-to-head experiment.

Table~\ref{table:PythonOpenSummary} summarizes the aging indicators observed in the Python Open applications. Memory usage presented the most consistent degradation evidence. All three applications showed statistically significant upward trends, indicating progressive resource accumulation during sustained execution. Image Converter presented the highest memory-growth slope and the highest average memory usage, suggesting that the image-processing workload imposed stronger memory pressure on the application under test. Uptime also showed a high memory-growth slope, followed by Monitor. The non-overlapping confidence intervals for memory usage indicate that the three applications exhibited distinct memory-growth rates.

\begin{table*}[!t]
\centering
\scriptsize
\caption{Summary of aging indicators for the Python Open applications. Superscript $^{\dagger}$ denotes statistically significant degradation.} 
\label{table:PythonOpenSummary}
\begin{tabular}{lcccccc}
\toprule
\textbf{Application} &
\textbf{Memory slope} &
\textbf{Mean memory} &
\textbf{RT slope} &
\textbf{Mean RT} &
\textbf{Throughput slope} &
\textbf{Mean TP} \\
&
\textbf{(GB/hour)} &
\textbf{(GB)} &
\textbf{(ms/hour)} &
\textbf{(ms)} &
\textbf{(req/s)} &
\textbf{(req/s)} \\
\midrule
Image Converter &
$8.239 \times 10^{-3\dagger}$ &
2.044 &
$-4.579 \times 10^{-2}$ &
6398.109 &
$-1.701 \times 10^{-4}$ &
1.565 \\

Monitor &
$3.255 \times 10^{-3\dagger}$ &
1.223 &
$7.842 \times 10^{-4\dagger}$ &
15.009 &
$-3.623 \times 10^{-3\dagger}$ &
66.311 \\

Uptime &
$5.043 \times 10^{-3\dagger}$ &
1.342 &
$4.112^{\dagger}$ &
171.754 &
$-2.860 \times 10^{-1\dagger}$ &
7.561 \\
\bottomrule
\end{tabular}
\end{table*}

Response time and throughput showed a more differentiated behavior. For response time, Monitor and Uptime presented statistically significant positive trends, while Image Converter did not show clear progressive latency degradation because its confidence interval crossed zero. The strongest latency degradation was observed in Uptime, with a Sen's slope of 4.112 ms/hour, indicating a substantial increase in response time over prolonged execution. For throughput, Monitor and Uptime also presented statistically significant negative trends, indicating progressive loss of processing capacity. Uptime again showed the strongest degradation, with a Sen's slope of $-2.860 \times 10^{-1}$ requests/s. Image Converter had the lowest average throughput, which is consistent with its heavier image-processing workload, but its confidence interval crossed zero, suggesting that it was throughput-limited on average without clear evidence of progressive throughput degradation.

Table~\ref{table:PythonOpenVsLLM} compares the Python Open applications with the corresponding Python LLM-generated implementations evaluated earlier in this paper. The comparison shows that the Python Open applications exhibited stronger memory growth in all three scenarios. For Image Converter, the memory-growth slope increased from $3.744 \times 10^{-3}$ in the LLM-generated implementation to $8.239 \times 10^{-3}$ in the Python Open implementation. For Monitor, the slope increased from $2.324 \times 10^{-3}$ to $3.255 \times 10^{-3}$, and for Uptime it increased from $2.810 \times 10^{-3}$ to $5.043 \times 10^{-3}$. This indicates that memory growth is not limited to LLM-generated systems and may also appear, sometimes more strongly, in human-written implementations.

\begin{table*}[!t]
\centering
\scriptsize
\caption{Comparison between Python LLM-generated and Python Open applications.}
%RT denotes response time. A degradation signal corresponds to a statistically significant positive trend for memory usage and response time, and to a statistically significant negative trend for throughput.}
\label{table:PythonOpenVsLLM}
\begin{tabular}{llcccccc}
\toprule
\textbf{Application} &
\textbf{Implementation} &
\textbf{Memory slope} &
\textbf{Memory deg.} &
\textbf{RT slope} &
\textbf{RT deg.} &
\textbf{Throughput slope} &
\textbf{Throughput deg.} \\
\midrule

Image Converter &
LLM-gen. Python &
$3.744 \times 10^{-3}$ &
\checkmark &
$-6.88 \times 10^{-1}$ &
-- &
$4.22 \times 10^{-6}$ &
-- \\

&
Python Open &
$8.239 \times 10^{-3}$ &
\checkmark &
$-4.579 \times 10^{-2}$ &
-- &
$-1.701 \times 10^{-4}$ &
-- \\

\midrule

Monitor &
LLM-gen. Python &
$2.324 \times 10^{-3}$ &
\checkmark &
$3.49 \times 10^{-5}$ &
-- &
$4.09 \times 10^{-4}$ &
-- \\

&
Python Open &
$3.255 \times 10^{-3}$ &
\checkmark &
$7.842 \times 10^{-4}$ &
\checkmark &
$-3.623 \times 10^{-3}$ &
\checkmark \\

\midrule

Uptime &
LLM-gen. Python &
$2.810 \times 10^{-3}$ &
\checkmark &
$-3.00 \times 10^{-3}$ &
-- &
$5.54 \times 10^{-3}$ &
-- \\

&
Python Open &
$5.043 \times 10^{-3}$ &
\checkmark &
4.112 &
\checkmark &
$-2.860 \times 10^{-1}$ &
\checkmark \\

\bottomrule
\end{tabular}
\end{table*}

The comparison also shows that the Python LLM-generated implementations of Image Converter, Monitor, and Uptime did not present clear progressive degradation in response time or throughput, whereas the Python Open versions of Monitor and Uptime degraded across all three metrics. This difference is particularly evident for Uptime, where the Python Open implementation showed statistically significant memory growth, increasing response time, and declining throughput. The Image Converter comparison was more mixed. Although the Python Open implementation exhibited stronger memory growth than the LLM-generated version, it did not show clear latency or throughput degradation. This suggests that the aging behavior of human-written systems may be strongly influenced by concrete implementation choices, libraries, persistence mechanisms, and workload characteristics.

% These results show that software aging symptoms are not exclusive to LLM-generated applications. In the evaluated scenarios, the Python Open Monitor and Uptime applications presented stronger overall aging evidence than their Python LLM-generated counterparts, while Image Converter showed stronger memory growth but no clear performance degradation. Therefore, RQ4 indicates that human-written implementations can exhibit aging trends that are comparable to, or stronger than, those observed in LLM-generated implementations. However, because the systems are functionally related but not exact replicas, this result should be interpreted as contextual evidence rather than as a controlled human-versus-LLM comparison.
These results show that the software aging symptoms observed in LLM-generated systems align with degradation patterns that can also occur in human-written implementations. In the evaluated scenarios, the Python Open Monitor and Uptime applications presented stronger overall aging evidence than their Python LLM-generated counterparts, while Image Converter showed stronger memory growth but no clear performance degradation. Therefore, RQ4 indicates that the aging behavior observed in LLM-generated applications is not an isolated phenomenon, but part of a broader class of long-running software degradation effects that may also affect manually developed systems. However, because the systems are functionally related but not exact replicas, this result should be interpreted as contextual evidence rather than as a controlled human-versus-LLM comparison.

\begin{tcolorbox}[
breakable,
colback=gray!3,
colframe=gray!40,
boxrule=0.5pt,
arc=1pt,
left=4pt,
right=4pt,
top=4pt,
bottom=4pt,
title=\textbf{Answer to RQ4},
fonttitle=\bfseries
]
The exploratory comparison shows that the software aging symptoms observed in LLM-generated systems align with degradation patterns also found in human-written implementations. The human-written Python open-source implementations exhibited statistically significant memory growth in all analyzed scenarios, and the Monitor and Uptime applications also showed degradation in response time and throughput. Compared with the corresponding Python LLM-generated implementations, the Python Open applications presented stronger memory growth in all three scenarios and stronger overall aging evidence in Monitor and Uptime. The Image Converter comparison was more mixed, since the Python Open implementation showed stronger memory growth but no clear latency or throughput degradation. Therefore, RQ4 indicates that LLM-generated and human-written implementations can exhibit comparable aging trends, although the human-written systems showed stronger degradation in some evaluated scenarios. However, because the systems are functionally related but not exact replicas, this result should be interpreted as contextual evidence rather than as a controlled human-versus-LLM comparison.
\end{tcolorbox}

\subsection{Threats to validity}

Threats to the validity of this work are described below.

\begin{itemize}

\item \textbf{Attribution of degradation causes.} This study evaluates complete backend application environments, including generated code, frameworks, runtimes, external tools, libraries, operating-system resources, and LLM-based generation platforms. This design reflects how generated applications are executed in practice, but it limits the ability to attribute degradation symptoms to a single factor. Therefore, the results should be interpreted as evidence of aging symptoms in complete LLM-generated application environments, rather than as proof that degradation is caused only by the application code or by the programming language.

\item \textbf{Static-analysis interpretation.} The static analysis combined traditional tools with an LLM-assisted code-review workflow to identify plausible aging-related mechanisms. However, static analysis cannot determine whether a potential mechanism was actually activated during the long-duration experiments. Some findings depend on specific runtime conditions, such as unique input values, cleanup failures, external command hangs, or request cancellation. In addition, LLM-assisted review may introduce false positives or miss implementation details. To mitigate this threat, the findings were filtered using explicit aging-related criteria and interpreted together with the runtime trends. Thus, the static-analysis results should be treated as explanatory evidence, not as definitive proof of causality.

\item \textbf{Experimental scope and workload.} The experiment included four service-oriented applications derived from BaxBench~\cite{Vero2025BaxBench:}, covering image processing, file handling, structured data manipulation, process inspection, and network availability checking. However, these scenarios do not represent all types of generated software. The experiments also used a fixed workload configuration with 10 concurrent threads, a request dispatch interval of 0.1 seconds, and 48 hours of continuous execution. This supports controlled comparison across application-environment combinations, but different request rates, concurrency levels, input sizes, failure conditions, longer executions, or repeated independent runs may expose different aging patterns.

\item \textbf{Monitoring granularity.} Memory consumption was monitored at the operating-system level rather than at the individual application-process level. This captures the resource pressure experienced by the server environment during long-running execution, but it may also include effects from the operating system, runtime libraries, background activity, or auxiliary processes. Therefore, memory trends should be interpreted as evidence of memory-related aging symptoms at the server-environment level, not as definitive proof of application-level memory leaks.

\item \textbf{Human-written comparison.} The comparison with human-written implementations was exploratory. The selected open-source systems implement scenarios that are functionally related to the evaluated LLM-generated applications, but they are not exact replicas. They may differ in frameworks, libraries, persistence mechanisms, architecture, code organization, and implementation choices. Therefore, the comparison should not be interpreted as a controlled human-versus-LLM experiment. Its purpose is to contextualize whether similar aging symptoms also appear in manually developed systems under comparable workload, monitoring, and statistical-analysis procedures.

\end{itemize}

\section{Conclusion}
\label{conclusion}

This paper investigated software aging symptoms in applications automatically generated by LLMs. Through a controlled experimental setup, four service-based applications implemented in JavaScript, Python, and Rust were subjected to long-duration workloads, and their behavior was monitored over 48 hours using memory usage, response time, and throughput as primary indicators. The results show that memory consumption was the most consistent indicator of potential software aging, with statistically significant upward trends in most application-language combinations. Response time and throughput presented more heterogeneous behavior, with degradation appearing only in specific configurations and sometimes being influenced by transient performance anomalies. These findings indicate that functional correctness alone is insufficient to assess the operational reliability of LLM-generated applications intended to run continuously.

The cross-environment analysis showed that aging symptoms vary according to the interaction among application scenario, programming language, backend framework, runtime environment, and generation platform. Python presented the highest memory growth rates in three of the four evaluated applications, namely Credit Card, Monitor, and Uptime, while JavaScript presented the highest memory growth in the Image Converter scenario. However, the results should not be interpreted as an isolated language comparison, since each environment combines a specific language, framework, runtime, and generation platform. This reinforces the need to evaluate LLM-generated systems as complete generation-and-execution environments.

The static analysis provided complementary evidence for interpreting the runtime results. The clearest potential aging mechanisms were associated with persistent application or database state without deletion, size limits, or expiration policies, especially in the Credit Card and Uptime scenarios. Temporary-resource retention and conditional subprocess retention were also identified in the Image Converter scenario. These findings helped explain several degradation trends observed during execution, but the correspondence was not uniform across all configurations. Some mechanisms depended on specific runtime conditions that may not have been sufficiently activated by the experimental workload, while some runtime degradation trends did not have a direct application-level mechanism visible in the source code. Thus, static analysis and long-duration runtime experimentation should be treated as complementary sources of evidence rather than interchangeable methods.

The exploratory comparison with human-written implementations showed that software aging symptoms are not exclusive to LLM-generated applications. In the evaluated scenarios, some human-written systems exhibited aging trends comparable to or stronger than those observed in the corresponding LLM-generated implementations. This suggests that aging behavior may depend more on concrete implementation choices, libraries, frameworks, persistence mechanisms, and resource-management decisions than on whether the code was produced by a human developer or generated by an LLM. However, because the human-written systems were not exact replicas of the generated applications, these findings should be interpreted as contextual evidence rather than as a controlled comparison between human and LLM-generated software.

Future work will focus on validating the identified mechanisms through targeted experiments, workload variation, failure injection, and finer-grained application-level monitoring of process memory, heap usage, file descriptors, subprocesses, database state, and open connections. We also plan to isolate the effects of language ecosystem, backend framework, and generation platform by controlling these factors independently. Finally, future studies should examine broader classes of applications, additional LLM-based development environments, and prompt patterns that may influence the long-term stability of generated software.

\section*{Acknowledgements}

The authors acknowledge the support of the CNPq, Brazil, through project number 401147/2025-8.

\bibliographystyle{IEEEtran}
\bibliography{full_bibliography}

@misc{linux_proc_docs,
  title        = {{The /proc Filesystem}},
  author       = {{The Linux Kernel Developers}},
  howpublished = {\url{https://www.kernel.org/doc/html/v6.15/filesystems/proc.html}},
  note         = {Accessed: Aug. 7, 2026}
}

@misc{psutil_docs,
  title        = {{psutil API Reference}},
  author       = {{psutil Developers}},
  howpublished = {\url{https://psutil.io/api/}},
  note         = {Accessed: Aug. 7, 2026}
}

@misc{nodejs_cluster,
  title        = {{Cluster}},
  author       = {{Node.js Contributors}},
  howpublished = {\url{https://nodejs.org/api/cluster.html}},
  note         = {Accessed: Aug. 7, 2026}
}

@article{MOURA2026112715,
title = {Machine learning for software aging detection: A systematic mapping study},
journal = {Journal of Systems and Software},
volume = {234},
pages = {112715},
year = {2026},
issn = {0164-1212},
doi = {https://doi.org/10.1016/j.jss.2025.112715},
url = {https://www.sciencedirect.com/science/article/pii/S016412122500384X},
author = {Rafael José Moura and Maria Gizele Nascimento and Fumio Machida and Domenico Cotroneo and Ermeson Andrade}
}

@inproceedings{machida2013effectiveness,
  title={On the effectiveness of Mann-Kendall test for detection of software aging},
  author={Machida, Fumio and Andrzejak, Artur and Matias, Rivalino and Vicente, Elder},
  booktitle={2013 IEEE International Symposium on Software Reliability Engineering Workshops (ISSREW)},
  pages={269--274},
  year={2013},
  organization={IEEE}
}

@article{couto2024comparative,
  title={A Comparative Analysis of Software Aging in Relational Database System Environments},
  author={Couto, Herderson and Machida, Fumio and Callou, Gustavo and Andrade, Ermeson},
  journal={IEEE Transactions on Emerging Topics in Computing},
  year={2024},
  publisher={IEEE}
}

@inproceedings{pietrantuono2022empirical,
  title={An empirical study on software aging of long-running object detection algorithms},
  author={Pietrantuono, Roberto and Cotroneo, Domenico and Andrade, Ermeson and Machida, Fumio},
  booktitle={2022 IEEE 22nd International Conference on Software Quality, Reliability and Security (QRS)},
  pages={1091--1102},
  year={2022},
  organization={IEEE}
}

@inproceedings{andrade2021memory,
  title={Memory Degradation Analysis in Private and Public Cloud Environments},
  author={Andrade, Ermeson and Machida, Fumio and Pietrantuono, Roberto and Cotroneo, Domenico},
  booktitle={2021 IEEE International Symposium on Software Reliability Engineering Workshops (ISSREW)},
  pages={33--39},
  year={2021},
  organization={IEEE}
}

@article{cotroneo:2014,
  title={A survey of software aging and rejuvenation studies},
  author={Cotroneo, Domenico and Natella, Roberto and Pietrantuono, Roberto and Russo, Stefano},
  journal={ACM Journal on Emerging Technologies in Computing Systems (JETC)},
  volume={10},
  number={1},
  pages={1--34},
  year={2014},
  publisher={Acm New York, NY, USA}
}

@inproceedings{valentim2016systematic,
  title={A systematic mapping review of the first 20 years of software aging and rejuvenation research},
  author={Valentim, Nathalia Assis and Macedo, Autran and Matias, Rivalino},
  booktitle={IEEE International Symposium on Software Reliability Engineering Workshops (ISSREW)},
  pages={57--63},
  year={2016},
  organization={IEEE}
}

@INPROCEEDINGS{Parnas1994,
  author={Parnas, D.L.},
  booktitle={16th International Conference on Software Engineering}, 
  title={Software aging}, 
  year={1994},
  volume={},
  number={},
  pages={279-287},
  doi={10.1109/ICSE.1994.296790}}

@article{mann1945nonparametric,
  title={Nonparametric tests against trend},
  author={Mann, Henry B},
  journal={Econometrica: Journal of the econometric society},
  pages={245--259},
  year={1945},
  publisher={JSTOR}
}

@article{sen1968estimates,
  title={Estimates of the regression coefficient based on Kendall's tau},
  author={Sen, Pranab Kumar},
  journal={Journal of the American statistical association},
  volume={63},
  number={324},
  pages={1379--1389},
  year={1968},
  publisher={Taylor \& Francis}
}

@INPROCEEDINGS{Dias,
  author={Dias, Douglas and Machida, Fumio and Andrade, Ermeson},
  booktitle={2022 IEEE International Symposium on Software Reliability Engineering Workshops (ISSREW)}, 
  title={Analysis of Software Aging in a Blockchain Platform}, 
  year={2022},
  volume={},
  number={},
  pages={170-177},
  doi={10.1109/ISSREW55968.2022.00064}}

@INPROCEEDINGS{imageCloudEdge,
  author={Andrade, Ermeson and Machida, Fumio and Pietrantuono, Roberto and Cotroneo, Domenico},
  booktitle={2020 IEEE International Symposium on Software Reliability Engineering Workshops (ISSREW)}, 
  title={Software Aging in Image Classification Systems on Cloud and Edge}, 
  year={2020},
  volume={},
  number={},
  pages={342-348},
  doi={10.1109/ISSREW51248.2020.00099}}

@article{Lyu2024Automatic,title={Automatic Programming: Large Language Models and Beyond},author={Michael R. Lyu and Baishakhi Ray and Abhik Roychoudhury and Shin Hwei Tan and Patanamon Thongtanunam},journal={ACM Transactions on Software Engineering and Methodology},year={2024},volume={34},pages={1 - 33},doi={10.1145/3708519}}

@article{Grebenshchikov2012Synthesizing,title={Synthesizing software verifiers from proof rules},author={S. Grebenshchikov and Nuno P. Lopes and C. Popeea and A. Rybalchenko},journal={Proceedings of the 33rd ACM SIGPLAN Conference on Programming Language Design and Implementation},year={2012},doi={10.1145/2254064.2254112}}

@article{Balzer1985A,title={A 15 Year Perspective on Automatic Programming},author={R. Balzer},journal={IEEE Transactions on Software Engineering},year={1985},volume={SE-11},pages={1257-1268},doi={10.1109/TSE.1985.231877}}

@article{Fan2022Improving,title={Improving automatically generated code from Codex via Automated Program Repair},author={Zhiyu Fan and Xiang Gao and Abhik Roychoudhury and Shin Hwei Tan},journal={ArXiv},year={2022},volume={abs/2205.10583},doi={10.48550/arXiv.2205.10583}}

@article{Matias2010Accelerated,title={Accelerated Degradation Tests Applied to Software Aging Experiments},author={Rivalino Matias and P. A. Barbetta and Kishor S. Trivedi and Paulo José de Freitas Filho},journal={IEEE Transactions on Reliability},year={2010},volume={59},pages={102-114},doi={10.1109/TR.2009.2034292}}

@article{Grottke2006Analysis,title={Analysis of Software Aging in a Web Server},author={Michael Grottke and Lei Li and K. Vaidyanathan and Kishor S. Trivedi},journal={IEEE Transactions on Reliability},year={2006},volume={55},pages={411-420},doi={10.1109/TR.2006.879609}}

@article{Grottke2008The,title={The fundamentals of software aging},author={Michael Grottke and Rivalino Matias and Kishor S. Trivedi},journal={2008 IEEE International Conference on Software Reliability Engineering Workshops (ISSRE Wksp)},year={2008},pages={1-6},doi={10.1109/ISSREW.2008.5355512}}

@article{Vero2025BaxBench:,title={BaxBench: Can LLMs Generate Correct and Secure Backends?},author={Mark Vero and Niels Mündler and Victor Chibotaru and Veselin Raychev and Maximilian Baader and Nikola Jovanovi'c and Jingxuan He and Martin T. Vechev},journal={ArXiv},year={2025},volume={abs/2502.11844},doi={10.48550/arXiv.2502.11844}}

@article{2024Accelerating,
title={Accelerating Software Development with Artificial Intelligence},
author={Anupriya and Paras Jain and Lipika Goel and R. Sharma and Sanjay Jasola and Amjad Ali},
journal={2024 International Conference on Artificial Intelligence and Emerging Technology (Global AI Summit)},
year={2024},
pages={727-732},
doi={10.1109/GlobalAISummit62156.2024.10947773}
}

@article{chauhan2025llm,
  title={Llm-generated microservice implementations from restful api definitions},
  author={Chauhan, Saurabh and Rasheed, Zeeshan and Sami, Abdul Malik and Zhang, Zheying and Rasku, Jussi and Kemell, Kai-Kristian and Abrahamsson, Pekka},
  journal={arXiv preprint arXiv:2502.09766},
  year={2025}
}

@article{Meem2024Exploring,title={Exploring Experiences with Automated Program Repair in Practice},author={Fairuz Nawer Meem and Justin Smith and Brittany Johnson},journal={2024 IEEE/ACM 46th International Conference on Software Engineering (ICSE)},year={2024},pages={1047-1057},doi={10.1145/3597503.3639182}}

@article{yang2024automated,
  title={Exploring and Unleashing the Power of Large Language Models in Automated Code Translation},
  author={Yang, Zhen and Liu, Fang and Yu, Zhongxing and Li, Jia and outros },
  journal={arXiv preprint arXiv:2404.14646},
  year={2024}
}

@inproceedings{SqlServerAgingNascimento2024,
author = {Nascimento, Maria Gizele and Moura, Rafael Jos\'{e} and Machida, Fumio and Andrade, Ermeson},
title = {Comparison of Machine Learning Algorithms for Detecting Software Aging in SQL Server},
year = {2024},
isbn = {9798400717406},
publisher = {Association for Computing Machinery},
address = {New York, NY, USA},
url = {https://doi.org/10.1145/3697090.3699798},
doi = {10.1145/3697090.3699798},
booktitle = {Proceedings of the 13th Latin-American Symposium on Dependable and Secure Computing},
pages = {159–164},
numpages = {6},
location = {
},
series = {LADC '24}
}

@INPROCEEDINGS{AgingTestFrameworkDias2025,
  author={Dias, Douglas and Machida, Fumio and Andrade, Ermeson},
  booktitle={2023 IEEE 34th International Symposium on Software Reliability Engineering Workshops (ISSREW)}, 
  title={Software Aging Analysis in a Testing Framework}, 
  year={2023},
  volume={},
  number={},
  pages={222-229},
  doi={10.1109/ISSREW60843.2023.00077},
  ISSN={},
  month={Oct},}

@misc{lu2025webgenbenchevaluatingllmsgenerating,
      title={WebGen-Bench: Evaluating LLMs on Generating Interactive and Functional Websites from Scratch}, 
      author={Zimu Lu and Yunqiao Yang and Houxing Ren and Haotian Hou and Han Xiao and Ke Wang and Weikang Shi and Aojun Zhou and Mingjie Zhan and Hongsheng Li},
      year={2025},
      eprint={2505.03733},
      archivePrefix={arXiv},
      primaryClass={cs.CL},
      url={https://arxiv.org/abs/2505.03733}, 
}

@inproceedings{costa2026case,
  title={A Case Study on Software Aging in LLM-Generated Python Applications},
  author={Costa, Gustavo and Santos, Cesar and Natella, Roberto and Andrade, Ermeson},
  booktitle={Workshop de Testes e Toler{\^a}ncia a Falhas (WTF)},
  pages={13--25},
  year={2026},
  organization={SBC}
}

@inproceedings{Santos_2025,
   title={Investigating Software Aging in LLM-Generated Software Systems},
   url={http://dx.doi.org/10.1109/ISSREW67781.2025.00090},
   DOI={10.1109/issrew67781.2025.00090},
   booktitle={2025 IEEE 36th International Symposium on Software Reliability Engineering Workshops (ISSREW)},
   publisher={IEEE},
   author={Santos, César and Andrade, Ermeson and Natella, Roberto},
   year={2025},
   month=Oct, pages={314–321} }

@article{Jamil2025Can,title={Can LLMs Generate Higher Quality Code Than Humans? An Empirical Study},author={M. Jamil and Shamsa Abid and S. Shamail},journal={2025 IEEE/ACM 22nd International Conference on Mining Software Repositories (MSR)},year={2025},pages={478-489},doi={10.1109/msr66628.2025.00081}}

@article{Molison2025Is,title={Is LLM-Generated Code More Maintainable \& Reliable Than Human-Written Code?},author={Alfred Santa Molison and Marcia Moraes and Glaucia Melo and Fabio Santos and Wesley K. G. Assunção},journal={2025 ACM/IEEE International Symposium on Empirical Software Engineering and Measurement (ESEM)},year={2025},pages={151-162},doi={10.1109/esem64174.2025.00036}}

@article{Li2025A,title={A Preliminary Study on the Robustness of Code Generation by Large Language Models},author={Zike Li and Mingwei Liu and Anji Li and Kaifeng He and Yanlin Wang and Xing Peng and Zibin Zheng},year={2025},doi={}}

@article{Kharma2025Security,title={Security and Quality in LLM-Generated Code: A Multi-Language, Multi-Model Analysis},author={Mohammed Kharma and Soohyeon Choi and Mohammed Alkhanafseh and David A. Mohaisen},journal={ArXiv},year={2025},volume={abs/2502.01853},doi={10.48550/arxiv.2502.01853}}

@article{Shehab2024Evaluating,title={Evaluating Large Language Models for Code Generation: Assessing Accuracy, Quality, and Performance},author={Mohammed A. Shehab and Mohammad Wardat and Safwan Omari and Yaser Jararweh},journal={2024 2nd International Conference on Foundation and Large Language Models (FLLM)},year={2024},pages={407-416},doi={10.1109/fllm63129.2024.10852439}}

@article{Tambon2024Bugs,title={Bugs in large language models generated code: an empirical study},author={Florian Tambon and Arghavan Moradi-Dakhel and Amin Nikanjam and F. Khomh and Michel C. Desmarais and G. Antoniol},journal={Empirical Software Engineering},year={2024},volume={30},doi={10.1007/s10664-025-10614-4}}

@article{Liu2024Beyond,title={Beyond Functional Correctness: Exploring Hallucinations in LLM-Generated Code},author={Fang Liu and Yang Liu and Lin Shi and Zhen Yang and Li Zhang and Xiaoli Lian and Zhongqi Li and Yuchi Ma},journal={IEEE Transactions on Software Engineering},year={2024},volume={52},pages={1037-1055},doi={10.1109/tse.2026.3657432}}

@misc{Peng2025CWEval,
      title={CWEval: Outcome-driven Evaluation on Functionality and Security of LLM Code Generation}, 
      author={Jinjun Peng and Leyi Cui and Kele Huang and Junfeng Yang and Baishakhi Ray},
      year={2025},
      eprint={2501.08200},
      archivePrefix={arXiv},
      primaryClass={cs.SE},
      url={https://arxiv.org/abs/2501.08200}, 
}

@article{cotroneo2020comprehensive,
  title={{A Comprehensive Study on Software Aging across Android Versions and Vendors}},
  author={Cotroneo, Domenico and Iannillo, Antonio Ken and Natella, Roberto and Pietrantuono, Roberto},
  journal={Empirical Software Engineering},
  volume={25},
  number={5},
  pages={3357--3395},
  year={2020},
  publisher={Springer}
}

\end{document}